\documentclass[12pt]{article}
\usepackage{setspace}
\usepackage{amsmath,amssymb,mathrsfs}
\usepackage{color}
\usepackage{hyperref}
\begin{document}
\setlength{\topmargin}{-1cm} 
\setlength{\oddsidemargin}{-0.25cm}
\setlength{\evensidemargin}{0cm}
\newcommand{\e}{\epsilon}
\newcommand{\beq}{\begin{equation}}
\newcommand{\eeql}[1]{\label{#1}\end{equation}}
\newcommand{\eeq}{\end{equation}}
\newcommand{\bea}{\begin{eqnarray}}
\newcommand{\eeal}[1]{\label{#1}\end{eqnarray}}
\newcommand{\eea}{\end{eqnarray}}
\renewcommand{\Im}{{\rm Im}\,}
\renewcommand{\Re}{{\rm Re}\,}
\newcommand{\diag}{{\rm diag} \, }
\newcommand{\Tr}{{\rm Tr}\,}
\def\draftnote#1{{\color{red} #1}}
\def\bldraft#1{{\color{blue} #1}}
\def\n{n \cdot v}
\def\ni{n\cdot v_I}
\begin{titlepage}
\begin{center}

\vskip 4 cm

{\Large \bf A Universal Feature for the Higgs Phenomenon in Anti de Sitter Space}

\vskip 1 cm

{Massimo Porrati$^a$ \footnote{E-mail: \href{mailto:mp9@nyu.edu}{mp9@nyu.edu}} and Alberto Zaffaroni$^{b,c}$
\footnote{Corresponding author. E-mail: \href{mailto:alberto.zaffaroni@mib.infn.it}{alberto.zaffaroni@mib.infn.it}}}

\vskip .75 cm

{\em a) Center for Cosmology and Particle Physics, \\ Department of Physics, New York University, \\ 726 Broadway, New York, NY 10003, USA}

\vskip .75cm

{\em b) Dipartimento di Fisica, Universit\`a di Milano-Bicocca, \\ Piazza della Scienza 3, 20126 Milano, Italy}

\vskip .75cm

{\em c) INFN, sezione di Milano-Bicocca,  \\ Piazza della Scienza 3, 20126 Milano, Italy}

\end{center}

\vskip 1.25 cm
\begin{abstract}
\noindent  
In Anti de Sitter space both massive and massless high-spin particles can have consistent local interactions. Both can couple
to conserved currents. In this paper we show that when the particles have spin one or greater, there exists a universal feature associated to the particle becoming massive: the currents possess a non-vanishing boundary flux.
 \end{abstract}
\end{titlepage}
\newpage

\section{Introduction}\label{intro}
Stanley Deser gave deep and long-lasting contributions to the understanding of the dynamics of high-spin particles. From
early no-go theorems on interacting massless spin $s>2$ particles~\cite{Aragone:1979hx} to the discovery that $s>1$ massless particle in 4-dimensional Anti de Sitter (AdS$_4$) space can propagate inside the light 
cone~\cite{Deser:1983mm} to the systematic analysis of massless and massive
high spin particles in de Sitter and AdS spaces~\cite{Deser:2001pe}, Stanley's impact has been a lasting one. In this paper, we add to Stanley's results on
high spin theories
by describing a general property common to the Higgs effect for particles of any spin $s\geq 1$ in AdS$_4$
spacetime. Our aim is to find an analog of a well known property of the Higgs mechanism in flat spacetime; 
namely, the
presence of massless poles corresponding to the propagation of lower-spin field in the two-point correlator of the current coupled to the high-spin field. 

The simplest example is
for spin $s=1$. Massless, CPT-preserving spin 1 representations contain only two helicities: $h=\pm1$. Massive 
representations contain also a zero-helicity state, which is provided by the Goldstone boson that is ``eaten'' by the spin 1.
The mass term in the propagator of the spin 1 state, represented by the vector field $W_\mu$, originates from the 2-point function of the conserved current $J_\mu$, which is coupled to $W_\mu$ through the standard interaction Lagrangian 
density $L_I=W_\mu J^\mu+..$. The $...$ denote non-minimal and quadratic terms in $W_\mu$ that are irrelevant for our
argument. In this language the Higgs mechanism is due the existence of a massless scalar field $\phi$ possessing a 
nonzero matrix element with $J_\mu$; namely $J_\mu=F \partial_\mu  \phi+..$. Here $F$ is a constant and 
again $...$ denote additional but 
inconsequential terms in $J_\mu$. Because of the Goldstone scalar $\phi$ contribution to the current, the two-point funtion of the current possesses
a zero-momentum pole
\beq
\int d^4 x e^{ipx} \langle J_\mu(x) J_\nu(0)\rangle = \frac{p_\mu p_\nu} {p^2} .
\eeql{intro1}
 The long-range $1/p^2$ pole in the two-point function of the current is the telltale of the Higgs mechanism. It is easy to 
 see that the presence of similar {\em Goldstone poles} is also necessary and,
 under appropriate conditions, sufficient for giving mass to massless particles
 of any integer spin $s>1$.  
  We could repeat this argument for half-integer spin particles starting with $s=3/2$ but we shall
 not, since it would be repetitious. Instead, we devote the rest of this paper to describing a feature of the Higgs mechanism in AdS$_4$
 space that is as general and sharp as the presence of a Goldstone pole in flat spacetime. This feature is the existence of
 a nonzero {\em boundary flux} for the conserved current $J_{\mu_1,..,\mu_s}$ that couples to the spin $s$ field
 $W_{\mu_1,..,\mu_s}$ as $L_I= W_{\mu_1,..,\mu_s} J^{\mu_1,..\mu_s}+...$. We review relevant group-theoretical 
 aspects of the Higgs mechanism in AdS$_4$ in the next section, which also explains why a nonzero boundary flux is
 a necessary and sufficient feature of the AdS$_4$ Higgs mechanism. Section 3 is instead dedicated to an explicit verification of this
 general feature in the case of spin $s=1,2$. 
 
\section{Higgs mechanism and boundary flux in AdS}\label{boundary}

Let us first review the group theoretical aspects of the Higgs mechanism in AdS$_4$.  Unitary and irreducible positive-energy representations of the isometry group $SO(2,3)$ are labeled by the minimum energy $E$ and the spin $s$ and are denoted by $D(E,s)$. A representation of integer spin $s\ge 1$ is unitary only in the range $E\ge s+1$ and the saturation of the bound corresponds to the massless case \cite{Fronsdal:1978vb,Breitenlohner:1982jf,Heidenreich:1980xi}.  In the saturation limit $E\rightarrow s+1$ the representation $D(E,s)$ becomes reducible, decomposing as
\beq
D(E,s) \rightarrow  D(s+1,s) \oplus D(s+2,s-1) \, ,\qquad E\rightarrow s+1 \, .\eeql{zaff1}
This is the representation-theoretic realization of the Higgs mechanism: a massless particle of spin $s$ becomes massive by ``eating'' a particle of spin $s-1$, the Goldstone boson.  
As we see from the decomposition~\eqref{zaff1}, the Goldstone mode is associated with the representation $D(s+2,s-1)$. In particular, a massless spin one becomes massive by eating a scalar $D(3,0)$ while a massless spin two needs to 
 eat a {\em massive} vector field $D(4,1)$. 

Consider now a spin-$s$ gauge field $W_s\equiv W_{\mu_1,\ldots ,\mu_s}$ and the corresponding conserved 
current it  couples to, $J_s\equiv J_{\mu_1,\ldots ,\mu_s}$. In a manifestly gauge invariant theory, $W_s$ can become massive  by eating a Golstone boson belonging to $D(s+2,s-1)$. This state can be either elementary or composite. In
any case, representation theory implies that the representation  $D(s+2,s-1)$ appears in the spectral decomposition of the self-energy for $W_s$. In flat space a composite Goldstone can only arise as an effect of strongly coupled dynamics. However, differently from flat space, the Goldstone mode can appear in AdS also at the perturbative level, so a spin $s$ field can acquire mass already through one-loop corrections.
This is due to the fact that  multi-particle states have a discrete energy spectrum in AdS and therefore even free fields can form bound states.

Examples of mass generation for gravity theory in AdS$_4$ coupled both to strongly and weakly interacting matter are
known. 
An example at strong coupling is given by the Karch-Randall (KR) brane model~\cite{Karch:2000ct}. This is  a general covariant theory that, from the 4d point of view, contain massive spin two fields but no massless graviton. The model has a holographic interpretation in terms of gravity in AdS$_4$ coupled to a strongly interacting CFT~\cite{Porrati:2001gx}.  
Unlike in other standard examples of holographic duality, here the four-dimensional physics is characterized by a transparent boundary condition at infinity that allows a flux of energy to escape from the boundary of AdS$_4$. A simple example of mass generation at weak coupling was given in~\cite{Porrati:2001db}, where it was shown that a graviton in AdS$_4$ becomes massive when it is coupled to a free conformal scalar. A crucial ingredient in the construction is again a suitable choice of boundary conditions for the scalar. A conformal scalar has two standard quantizations with a well defined and conserved energy that correspond to either a $D(1,0)$ or a $D(2,0)$ representation. Crucially, the graviton acquires a mass only when  boundary conditions that contain both modes are used, as for example the transparent boundary conditions discussed 
in~\cite{Avis:1977yn}.  When both representations are present there is a flux of energy at the boundary of AdS$_4$, and the theory makes sense only when coupled to a defect CFT$_3$ that can absorb the flux.

For spin one fields in AdS, a choice of charge-breaking boundary conditions at infinity can generate a mass for the gauge bosons at one-loop \cite{Porrati:2009dy,Rattazzi:2009ux}.
Even in this case there is a non zero flux at infinity, but this time charge, and not energy, is flowing through the boundary.

There are also examples of higher spin fields acquiring a mass through  a one-loop diagram of matter fields \cite{Girardello:2002pp}. Even in these examples the role of boundary conditions is crucial. 

In this paper we argue that the presence of a boundary flux of the  current $J_s\equiv J_{\mu_1,\ldots ,\mu_s}$ is in fact a universal feature of a spin $s\ge 1$ particle becoming massive in AdS$_4$, and therefore it uniquely characterizes  the presence of  a Goldstone boson, irrespective of it being a fundamental degree of freedom or a composite one.
For spin 1 currents the link between flux at infinity and spontaneous symmetry breaking was discussed in~\cite{ldp1,ldp2}. 
 In particular, a graviton in AdS$_4$ coupled to a generic CFT becomes massive if and only if   
there is  a boundary energy flux. 
This can be seen as follows. 

Euclidean AdS$_4$ is the hyperboloid in $\mathbb{R}^{1,4}$ defined by the quadratic equation
 \beq Y\cdot Y\equiv -Y_5^2+\sum_{k=1}^4 Y_k^2 =-L^2 \, .\eeq
 It is convenient to use coordinates $x^\mu=(z,x_i)$ with $i=1,2,3$
 \beq Y_4+Y_5 =\frac{L^2}{z}\, ,\quad   Y_4-Y_5= z+{1\over z}\sum_{i=1}^3 x^i,\quad Y_i=\frac{L x_i}{z} \, . \eeq
 Lorentzian AdS$_4$ is written in the same coordinates but with the replacement $x^3= i x^0$.
 In both cases we find the metric
 \beq d s^2= L^2 \, \frac{d z^2+d x_i d x^i}{z^2} \, ,\eeq
where the index $i$ runs over the three-dimensional slice and 
 the boundary is at $z=0$. 

Denoting with $\nabla_\mu$ the metric covariant derivative, the conservation law for $J_s$ is
\beq \nabla^\mu J_{\mu,\mu_1,\ldots ,\mu_{s-1}} =0 \, ,\eeq
It implies the following equation for the variation of the charge $Q(\tau)=\int_{t=\tau}\sqrt{|g|} g^{00}J_{0,\ldots ,0}$
\beq Q(\tau_2)-Q(\tau_1) =\int_{z=0} d^3 x \sqrt{|g|} g^{zz} J_{z,0, \ldots ,0} \, ,\eeq
where we integrate over the region $\tau_1\le t\le \tau_2$ and we assumed that the fields vanish sufficiently fast for large spatial coordinates $|x^i|$. Consider now  the integrated two-point
function of the current, contributing to the spin-$s$ self-energy
\beq 
\int_{z\rightarrow 0} d^3 x\sqrt{|\det g(z,x)|} g^{zz}(z,x) \langle J_{z,0, \ldots ,0}(z,x) J_{\mu_1,\ldots ,\mu_s} .(z^\prime,x^\prime)\rangle \eeq
We  make the point $(z,x)$ approach the boundary while $(z^\prime,x^\prime)$ is kept in the interior. In general, the two-point function  vanishes with some power of $z$ when $z\rightarrow 0$ and  we can have a non zero boundary integral only when it scales like $z^2$. 
 A generic mode $D(\Delta,s)$ appearing in the spectral decomposition of the two point function leads to a behavior
\beq\label{as} J_{z,\ldots ,z}(z,x)\sim z^\Delta\, ,\,\, \ldots \, , J_{z,i_1,\ldots ,i_{s-1}}(z,x)\sim z^{\Delta-s+1}\, ,\, \,  J_{i_1,\ldots ,i_{s}}(z,x)\sim z^{\Delta-s} \, ,\eeq
for $z\rightarrow 0$, when inserted in a correlation function. This can be proven by first noticing that under the 
general coordinate transformation $z\rightarrow z'=\lambda z$, $x^i\rightarrow x^{'\, i}=\lambda x^i$, the 4D 
components of a rank-$s$  tensor field transform as
\beq
h_{i_1,....,i_s}(z,x)\rightarrow h'_{i_1,....,i_s}(z',x')=\lambda^{-s} h_{i,....,i_s}(z,x).
\eeql{m2}
If the leading asymptotic term in $h_{i_1,....,_s}$ near $z=0$ is $h_{i_1,....,i_s}(z,x)\sim z^\beta \phi_{i_1,....,i_s}(x)$, eq.~\eqref{m2} implies that $\phi_{i_1,....,i_s}$ transforms as a conformal primary of weight $\Delta=\beta+s$:
\beq
\phi_{i_1,....,i_s}(x) \rightarrow \phi'_{i_1,....,i_s}(\lambda x)=\lambda^{-\beta-s} \phi_{i_1,....,i_s}(x).
\eeql{m3}
This identifies the constant $\Delta=\beta+s$  in the current $J_{i_1,....,i_s}$ as the conformal weight.
To identify the scaling of the other components of the current $J_{\mu_1,....,\mu_s}$ we use the homogeneous coordinate
technique developed in~\cite{Fronsdal:1978vb}. 
It was shown there that a tensor in $AdS_4$ is equivalent to a homogeneous 
tensor of (arbitrary) degree $N$ in the homogeneous coordinates $Y_M$ introduced above, which furthermore obeys the transversality condition $Y^Mh_{M,N_1,...,N_{s-1}}=0$.
The homogeneous tensors are written in terms of  the 4D ones using
\beq
h^{M_1,....,M_s}=\partial_{\mu_1}Y^{M_1}.... \partial_{\mu_s} Y^{M_s} h^{\mu_1,....,\mu_s} .
\eeql{m4}
In the limit $z\rightarrow 0$, all homogeneous coordinates are proportional to $z^{-1}$, so equating the left and right side 
of~\eqref{m4} for arbitrary numbers of indices $\mu_1=\mu_2=\ldots \mu_r=z$, $r\leq s$, we get the other equations in~\eqref{as}. 

The corresponding boundary flux
\beq \int_{z\rightarrow 0} d^3 x \sqrt{|g|} g^{zz} J_{z,0, \ldots ,0} \sim z^{\Delta-s-1} \, ,\eeq
vanishes for all irreducible massive modes with $\Delta>s+1$. However, when the spin $s$ field become massless, a Goldstone boson  appears in the spectral decomposition of its self energy. The current associated with such mode can be expressed as  $J_{\mu_1,\ldots , \mu_s}=\nabla_{(\mu_1} \phi_{\mu_2,\ldots , \mu_s)}$, where $\phi$ belongs to the $D(s+2,s-1)$ representation. With a gauge choice,
  the entire $J_{z,i_2,\ldots , i_s}$ component of the current at $\Delta=s+1$ can be written  as $\nabla_{(z} \phi_{i_2,\ldots , i_s)}$ in terms of the Goldstone mode. We then have a contribution
\beq J_{z,0,\ldots ,0}\sim\nabla_{z} \phi_{0,\ldots,0} +\ldots \sim z^2 \, ,\eeq
since $ \phi_{0,\ldots,0}\sim z^{\Delta-(s-1)}=z^3$ for $\Delta=s+2$. This contribution leads to a non-vanishing boundary flux. 

The previous argument shows that the presence of a Goldstone mode implies a boundary flux for $J_{z,0,\ldots ,0}$. The converse statement that a flux implies a Goldstone boson is slightly more complicated to formulate.
First of all, for the statement to be true we must consider only
  boundary conditions that preserve $SO(2,3)$. In this case the two point
  functions
  of currents possess a K\"all\'en-Lehmann spectral decomposition.
In the spectral decomposition of the  two point function of a spin $s$ current we can find modes  with all spins $0\le s^\prime \le s$~\cite{Meineri:2023mps}. They can be schematically written as
\beq \Pi_{(s-s^\prime)} \phi_{\mu_1,\ldots , \mu_{s^\prime}} \, , \eeql{proj}
where $\phi_{\mu_1,\ldots , \mu_{s^\prime}}$ belongs to the $D(\Delta,s^\prime)$ representation and $\Pi_{(s-s^\prime)}$ is a projector made with covariant derivatives and metric tensors. We considered previously the case $s^\prime=s$. Due to the conservation of the current, the only mode with $s^\prime=s-1$ that can appear in~\eqref{proj} is actually the Goldstone boson. On the other hand, there are plenty of modes with $s^\prime < s-1$ allowed by the conservation of the current that can contribute to the  two point function and, in principle, some of them  can lead to a boundary flux.\footnote{The same scaling argument that we used for modes with spin $s$ would lead generically to boundary fluxes of order $z^{\Delta-s}$ but now the unitary bound for $D(\Delta,s^\prime)$ only requires $\Delta\ge s^\prime +1$. It is possible that these terms are actually absent in the spectral decomposition. We will discuss explicitly this possibility for $s=2$ in the next section.} We can eliminate all these unwanted contributions if we consider the more general boundary flux 
\beq\label{gen}  \int_{z=0} d^3 x \sqrt{|g|} g^{zz} h^{i_1,\ldots ,i_{s-1}}J_{z, i_1, \ldots ,i_{s-1}} \, ,\eeq
where $h^{i_1\ldots i_{s-1}}$ is a symmetric tensor on the boundary, which is traceless and divergenceless with respect to 
the boundary metric, i.e. 
\beq \nabla_{i_1} h^{i_1,\ldots , i_{s-1}} =0 \, ,\qquad g_{i_1,i_2} h^{ i_1,i_2 \ldots ,i_{s-1}}=0 \, ,\eeq
where $i_k$ denote the boundary indices. Due to the form of the projectors $\Pi_{(s-s^\prime)}$, this generalized flux gets contributions only from intermediate states with $s^\prime=s$ or $s^\prime = s-1$ and therefore it is non-vanishing if and only if there exists a Goldstone boson. 

In the next sections we verify these statements for spin $s=1$ and $s=2$ by an explicit calculation using the AdS$_4$ K\"all\'en-Lehmann spectral representation of the two-point functions of  vector currents and the stress-energy tensor.

\section{K\"all\'en-Lehmann spectral representation in AdS$_4$}

The K\"all\'en-Lehmann spectral representation of two point functions of spin one and spin two conserved currents in AdS has been constructed explicitly in \cite{Osborn:1999az}.\footnote{The scalar case was originally considered in \cite{Dusedau:1985ue}. See~\cite{Meineri:2023mps} for a general discussion of symmetric, traceless but not necessarily conserved tensors.}  In this section we will investigate the relation between the occurrence of a Goldstone mode in the intermediate states and the boundary behavior of the two-point function. We need  some formalism first.

\subsection{Bi-tensors in AdS$_4$}
Following \cite{Allen:1985wd,Osborn:1999az}, we introduce the geodesic interval $\sigma(x,x^\prime)$ satisfying
\beq g^{\mu\nu}\partial_\mu\sigma \partial_\nu\sigma = 2\sigma \, ,\eeq
and the distance $\theta$ defined by $\sigma =\frac{L^2 \theta^2}{2}$, where $L$ is the AdS radius.

In AdS and more generally in a homogeneous space of constant curvature, any bi-tensor $T_{\mu\nu\ldots ;\alpha\beta\dots}(x,x^\prime)$ can be expanded in a basis formed by $g_{\mu\nu}$, $g_{\alpha\beta}$ and 
the derivatives $\partial_\mu\sigma$, $\partial_\alpha\sigma$ and $\partial_\mu\partial_\alpha\sigma$. Here and in the following $\mu, \nu, \ldots$ and $\alpha, \beta, \ldots$ are indices pertaining the tensorial properties at $x$ and  $x^\prime$, respectively, and $\partial_\mu \equiv \partial_{x_\mu}$ and similarly $\partial_\alpha \equiv \partial_{x^\prime_\alpha}$.

 We can also define the parallel transport from $x$ to $x^\prime$ along a geodesic through the bi-tensor $I^\mu_{\,\,\, \alpha}(x,x^\prime)$ defined by 
 \beq  \nabla_\sigma I^\mu_{\,\,\, \alpha}(x,x^\prime) =0 \, ,\qquad I^\mu_{\,\,\, \alpha}(x,x)=\delta^\mu_{\,\,\, \alpha} \, .\eeq
 The tensor $I_{\mu\alpha}$ can be combined with the metrics $g_{\mu\nu}$, $g_{\alpha\beta}$  and the unit vectors
 \beq \hat x_\mu = L \partial_\mu \theta\, ,\qquad \hat x^\prime_\alpha = L \partial_\alpha \theta\, ,\qquad \hat x_\mu I^\mu_{\,\,\, \alpha}=- \hat x^\prime_\alpha\, , \qquad I_\mu^{\,\,\, \alpha} \hat x^\prime_\alpha=-\hat x_\mu \, ,\eeq
 to form a basis.

 The geodesic distance between two points $Y$ and $Y^\prime$ with coordinates $(z,x_i)$ and $(z^\prime,x^\prime_i)$
 can be expressed in terms of \cite{Osborn:1999az}
 \beq u=\frac{(Y-Y^\prime)^2}{2L^2} =\frac{(z-z^\prime)^2+(x_i-x_i^\prime)(x^i-x^{\prime\, i})}{2 z z^\prime} \, ,\eeq
 as
 \beq\theta= {\rm arcosh} (u+1) \, .\eeq
 The parallel transport tensor is instead given by  
 \beq I_{\mu\alpha} =-\frac{\sinh \theta}{\theta} \partial_\mu\partial_\alpha \sigma -\frac{1-\sinh \theta/\theta}{L^2 \theta^2}  \partial_\mu \sigma\partial_\alpha \sigma \, .\eeq
 
 A function that will be important in the following is~\cite{Osborn:1999az}
 \beq G_{\Delta,s}(\theta) =\frac{1}{2 \pi^{3/2}L^{2s+2}}\frac{\Gamma(\Delta+s)}{\Gamma(\Delta-1/2)} e^{-(\Delta+s)\theta} F(\Delta+s,s+\frac32,\Delta-\frac12; e^{-2\theta}) \, .\eeq
 It is associated with a field of energy $\Delta$ and spin $s$. For $s=0$ it is precisely the Green function for the scalar operator 
 \beq  -\nabla^2 +\frac{\Delta(\Delta  -3)}{L^2} \, \, .\eeq
  If we fix $x^\prime$ to be a generic point in AdS$_4$ and send $x$ to the boundary, $z\rightarrow 0$, we find $\theta \sim -\log z$ and $G_{\Delta,s}\sim z^{\Delta+s}$. 
 
 \subsection{Spectral representation for currents}
 Consider now the K\"all\'en-Lehmann spectral representation for conserved currents in AdS$_4$. 
The two point function of currents has the general tensorial structure
\beq \langle J_\mu(x) J_\alpha (x^\prime) \rangle = \hat x_\mu \hat x_\alpha^\prime D(\theta) +\hat I_{\mu\alpha} E(\theta)\, ,\eeq
where for convenience we have defined
\beq \hat I_{\mu\alpha} =I_{\mu\alpha}+\hat x_\mu \hat x_\alpha^\prime \, .\eeq
The conservation of the current requires
\beq\label{deg1} \sinh \theta D^\prime(\theta)+3 \cosh \theta D(\theta) =-3 E(\theta) \, .\eeq
The two-point function is determined by a single function $D(\theta)$ whose spectral decomposition reads  \cite{Osborn:1999az}
\beq D(\theta) =\int_2^\infty d\Delta \rho(\Delta) G_{\Delta,1}(\theta) \, ,\eeq
where the lower integration limit is fixed by the unitary bound for spin one fields.
 
We now  fix $x^\prime$ to be a generic point in the bulk and send $x$ to the boundary by taking $z\rightarrow 0$. In this limit, an explicit computation shows that all components of $ \hat I_{\mu\alpha}$, $\hat x_\mu$ and  $\hat x_\alpha^\prime$ are $O(1)$ for $z\rightarrow 0$ with the exception of $\hat x_z \sim -\frac{L}{z}+O(z)$ and $\hat I_{i\alpha} =O(\frac{1}{z})$. Using $G_{\Delta,1}(\theta)\sim z^{\Delta+1}$, we see that a generic mode $\Delta>2$ in the spectral decomposition leads to $D(\theta)\sim z^{\Delta+1}$. Therefore $E(\theta)\sim z^\Delta$ and
\beq \langle J_z(x)J_\alpha(x^\prime)\rangle \sim z^\Delta \, \qquad \langle J_i(x)J_\alpha(x^\prime)\rangle \sim z^{\Delta-1}\, .\eeq
This result is in agreement with \eqref{as} and implies that there is no boundary charge flow  for all massive modes. 
The case $\Delta=2$, however, is special.  For example we see from \eqref{deg1}  that now $E(\theta)\sim z^4$ vanishes faster than $D(\theta)\sim z^2$. This  is a signal of the reducibility of the representation $D(\Delta,1)$ at the unitary bound
$\Delta=2$. The current can be written in terms of a scalar Goldstone boson $\phi$, transforming in the representation $D(3,0)$. We can explicitly write 
\beq \label{VEV} \langle J_\mu(x)J_\alpha(x^\prime)\rangle =\partial_\mu\partial_\alpha\langle \phi(x) \phi(x^\prime)\rangle \, ,\eeq
which is easily seen to vanish as $z^2$ for $\mu=z$ and $z^3$ for $\mu=i$ by taking into account that the propagator of $\phi$ is $G_{3,0}(\theta)\sim z^3$. 
Thus the presence of a Goldstone boson in the spectral decomposition leads to a boundary flux.

When a $U(1)$ gauge symmetry is broken by the VEV of a bulk scalar $\Phi$, the $s=0$ field in \eqref{VEV} is the standard elementary Goldstone boson, i.e. the phase $\phi$ of the scalar $\Phi=\rho e^{i \phi}$.

This elementary example shows that our analysis captures the lowest order effect in a perturbative expansion in the gauge coupling constant. The expansion is particularly effective for higher spin fields, which couple to spin-s current through irrelevant operators. In this case the validity of the perturbative expansion itself defines the regime of validity of the effective field theory.

\subsection{Spectral representation for the stress-energy tensor}
Consider now the K\"all\'en-Lehmann spectral representation  for the stress-energy tensor in AdS$_4$. 
The two point function  has the general tensorial structure
\bea\label{ka2} \langle T_{\mu\nu}(x) T_{\alpha\beta} (x^\prime) \rangle &=& \hat x_\mu \hat x_\nu \hat x_\alpha^\prime \hat x_\beta^\prime R(\theta) + (I_{\mu\alpha} \hat x_\nu  \hat x_\beta^\prime+{\rm symm}) S(\theta) 
+(I_{\mu\alpha} I_{\nu\beta}+I_{\nu\alpha} I_{\mu\beta}) T(\theta) \nonumber \\&+& (\hat x_\mu \hat x_\nu  g_{\alpha\beta} + \hat x_\alpha^\prime \hat x_\beta^\prime g_{\mu\nu}) U(\theta) +g_{\mu\nu} g_{\alpha\beta} V(\theta) \, .\eea
For convenience we also define
\beq Q = 2 T +\frac34 (R-4S)\, .\eeq

The conservation of the stress-energy tensors requires \cite{Osborn:1999az}
\bea\label{cons} -8 \, {\rm cosech}\, \theta (S-T) &=& Q^\prime + 4 \coth \theta Q \nonumber \\
-\frac{{\rm cosech}\, \theta}{3} (Q+10 T) &=&(S-T)^\prime + 4 \coth \theta (S-T)\, ,\eea
and the tracelessness of $T_{\mu\nu}$  imposes
\beq\label{trace}  R-4 S+4U = 2 T+U+4 V =0\, .\eeq

The K\"all\'en-Lehmann spectral representation for a traceless $T_{\mu\nu}$  reduces to \cite{Osborn:1999az}
\beq\label{KL2} Q(\theta) =\int_3^\infty d\Delta \rho(\Delta) G_{\Delta,2}(\theta) \, .\eeq

For a massive mode $\Delta>3$ we have $Q\sim z^{\Delta+2}$ and, from the conservation and tracelessness conditions \eqref{cons} and \eqref{trace}, we learn that all functions $R,S,T,U,V$ are generically of order $z^\Delta$ with the particular combinations $S-T\sim z^{\Delta+1}$ and $U+V\sim z^{\Delta+2}$. By re-organizing the terms in the two point function we can write
\bea \langle T_{\mu\nu}(x) T_{\alpha\beta} (x^\prime) \rangle =& - &3 \hat x_\mu \hat x_\nu \hat x_\alpha^\prime \hat x_\beta^\prime (U+V) + (\hat I_{\mu\alpha} \hat x_\nu  \hat x_\beta^\prime+{\rm symm}) (S-T)  \nonumber \\&+&
(\hat I_{\mu\alpha} \hat I_{\nu\beta}+\hat I_{\nu\alpha} \hat I_{\mu\beta}) T +(\hat x_\mu \hat x_\nu  (g_{\alpha\beta}- \hat x_\alpha^\prime \hat x_\beta^\prime) + \hat x_\alpha^\prime \hat x_\beta^\prime (g_{\mu\nu}-\hat x_\mu \hat x_\nu)) (U+V)  \nonumber \\&+&(g_{\mu\nu}-\hat x_\mu \hat x_\nu) (g_{\alpha\beta}- \hat x_\alpha^\prime \hat x_\beta^\prime) V \, ,\eea
and we see that the near-boundary behavior is
\beq \langle T_{zz}T_{\alpha\beta}\rangle \sim z^\Delta\, ,\qquad\langle T_{zi}T_{\alpha\beta}\rangle \sim z^{\Delta-1}\, ,\qquad\langle T_{ij}T_{\alpha\beta}\rangle \sim z^{\Delta-2}\, .\eeq
 This is in agreement with \eqref{as}; in particular, $\langle T_{z0}T_{\alpha\beta}\rangle \sim z^{\Delta-1}$  and therefore massive modes with $\Delta>3$ do no contribute to the flux of energy at infinity.  On the other hand, we can have a flux if $\Delta=3$, which corresponds to a reducible representation, as it can be seen from  conditions \eqref{cons} 
 and \eqref{trace}, which now predict  $S-T\sim z^6$ and $Q+10T\sim z^7$. In this case, the stress-energy tensor can be written in terms of a vector Goldstone boson $J_\mu$, transforming in the representation $D(4,1)$,
 \beq \langle T_{\mu\nu}(x) T_{\alpha\beta} (x^\prime) \rangle = \nabla_{(\mu} \langle J_{\nu )}(x) J_{(\alpha}(x^\prime) \rangle \overleftarrow{\nabla}_{\beta )}\, .\eeq
 
 There can be also spin zero modes in the spectral decomposition. These are associated with the following 
 possible tensorial structure
 \beq \langle T_{\mu\nu}(x) T_{\alpha\beta} (x^\prime) \rangle =\left(\nabla_\mu
\nabla_\nu -g_{\mu\nu}\nabla^2 +\frac{3}{L^2} g_{\mu\nu}\right ) F_0(\theta) \left(\overleftarrow{\nabla}_{\alpha} \overleftarrow{\nabla}_{\beta} -g_{\alpha\beta} \overleftarrow{\nabla}^2 +\frac{3}{L^2} g_{\alpha\beta}\right)\, .\eeq
 Notice that the spin zero part is transverse
 \beq \nabla^\mu \left(\nabla_\mu
\nabla_\nu -g_{\mu\nu}\nabla^2 +\frac{3}{L^2} g_{\mu\nu}\right ) = 0 \, , \eeq
but not traceless in general, since tracelessness implies
\beq 
\left(\nabla^2-\frac{4}{L^2}\right ) F_0 =0 ,
\eeq
which identifies a mode with $\Delta=4$.\footnote{Using the well-known relation between mass and dimension for a scalar in AdS$_4$, $m^2 L^2=\Delta(\Delta-3)$.
Being conserved and traceless, this particular spin zero mode can be formally included in  the spin two channel representation \eqref{KL2} by adding a mode $G_{2,2}(\theta)$ below the unitary bound \cite{Osborn:1999az}.}   All the effects of a non-traceless $T_{\mu\nu}$, including the Weyl anomaly,  can be re-absorbed in this tensorial structure.

The spectral decomposition now reads
\beq F_0(\theta) =\int_{\Delta_0}^\infty d\Delta \rho(\Delta) G_{\Delta,0}(\theta) \, ,\eeq
where the lowest limit of integration will be discussed in a moment. Since $G_{\Delta,0}\sim z^\Delta$ we see immediately that 
$\langle T_{z0}T_{\alpha\beta}\rangle \sim z^{\Delta-1}$. A natural choice for $\Delta_0$ would be the unitary bound for spin zero, $\Delta_0=\frac 12$.
In this case we see that we can have a non-zero flux and even divergent contributions from the spin zero modes. These contributions can be eliminated by considering a generalized flux of the form \eqref{gen} as discussed in section \ref{boundary}.

It was conjectured in \cite{Cappelli:1990yc,Osborn:1999az} that the correct lower bound for reasonable theories is actually $\Delta_0=4$. This bound ensures that the operator $\int dx^4 \sqrt{|g|} g^{\mu\nu} T_{\mu\nu}$ is well-defined. If this conjecture is true the spin zero modes would have $\Delta\ge4$ and there would be 
 no boundary flux at all for $T_{z0}$. It would be interesting to explore whether a similar argument could work also for higher spins.

\subsection*{Acknowledgements} 

   We thank Lorenzo di Pietro, Juan Maldacena and Marco Meineri for useful discussions and correspondence. 
   M.P. is supported in part by NSF grant PHY-2210349 and was supported by the Leverhulme Trust through a Leverhulme Visiting Professorship at
 Imperial College, London during the completion of this paper. A.Z. is partially supported by the INFN and by the MUR-PRIN grant No. 2022NY2MXY. A.Z. would like to thank The Center for Cosmology and Particle Physics at NYU for hospitality during the completion of this work.

\bibliographystyle{ytphys}
\bibliography{higgs} 
 
\end{document}